\documentclass[12pt]{article}
\usepackage[english]{babel}
\usepackage{a4wide}
\date{}
\begin{document}

\title{\bf {\large{Theorem of Levinson Via The  Spectral Density
\footnote{Presented by L.J.B. at the session on ``Semigroups, time
asymmetry and resonances'', within the XXV ICGTMP, Cocoyoc
(Mexico), August 2004.} }}}
\author{\normalsize{Luis J. Boya \footnote{luisjo@unizar.es}} \\
\normalsize{and} \\
\normalsize{Javier Casahorr\'an \footnote{javierc@unizar.es}} \\  \\ \\
\normalsize{Departamento de F\'{\i}sica Te\'orica} \\
\normalsize{Universidad de Zaragoza} \\
\normalsize{E-50009 Zaragoza} \\
\normalsize{SPAIN}}

\maketitle

\begin{abstract}

We deduce Levinson\'{}s theorem in non-relativistic quantum
mechanics in one dimension as a sum rule for the spectral density
constructed from asymptotic  data. We assume a self-adjoint
hamiltonian which guarantees completeness; the potential needs not
to be isotropic and a zero-energy resonance is automatically taken
into account. Peculiarities of this one-dimension case are
explained because of the ``critical'' character of the free case
$u(x) = 0$, in the sense that any atractive potential forms at
least a bound state. We believe this method is more general and
direct than the usual one in which one proves the theorem first
for single wave modes and performs analytical continuation.
\bigskip


\end{abstract}

\vfill \eject

\section{\large{Introduction.}}

Two generic results in potential scattering stand on their own,
and hold with wide generality. The first, the \emph{optical
theorem} \cite{Fe} stems from the fact that the scattered
``matter'' is taken away from the incoming wave, and hence the
scattering center ``casts a shadow'' in the forward direction as
to produce negative interference with the incoming beam; therefore
a relation must exist between the total scattering cross section
and the forward scattering amplitude. Originally proven in partial
waves for $3D$ scattering, the theorem holds with much more
generality; a simple but very general proof is offered in
\cite{Boy}. The theorem can be seen as a result of the
``completeness relation'' in ordinary space (also called
orthogonality), at a given energy. \par

The second result is \emph{Levinson\'{}s theorem}, which in a way
can be seen also as a consequence of completeness in
\emph{momentum space}. In its primitive form of 1949 Levinson\'{}s
theorem reads \cite{Lv1}

\begin{equation}\label{eq:1}
    n_{\ell} = (1/\pi) (\delta_{\ell}(0) - \delta_{\ell}(\infty))
\end{equation}

\noindent for the number $n_{\ell}$ of bound states of angular
momentum $\ell$ in a generic central potential $u(\vec{r}) =
u(|\vec{r}|)$ which produces a phase shift $\delta_{\ell}(k)$ for
scattering with energy $E = k^{2}$; the proof came up as a
byproduct of studies on uniqueness of potentials with a given
phase shift \cite{ Lv2}. Formula (\ref{eq:1}) was rediscovered in
1956 \cite{dW}. In 1957 Jauch \cite{Jau} established the theorem
as a consequence of the completeness relation for scattering
states and set it up in the general frame of operator theory. \par

The philosophy was that, if a potential generates bound states,
there should be a relation between them and scattering states as
the completeness relation has to be ``shared'' among them. Indeed,
Levinson\'{}s theorem is the only relation between bound states
and scattering states as it can be deduced from inverse scattering
theory. Since the work of Jauch, many studies followed and we
mention in particular the elementary deduction by Wellner
\cite{Wel} for s-waves in 3D, which we shall generalize in this
paper and the later studies of Newton relating the theorem to the
inverse problem in 1, 2 or 3 dimensions \cite{Nw1}, \cite{Nw2} .
\par

It is considerably more difficult to prove Levinson\'{}s theorem
than the optical theorem, although both share complementary
physical foundations. In this paper we prove the theorem as a sum
rule for the spectral density, which we take as more fundamental
entity; we shall work in one dimension with local potential where
all the features of the problem already show up. Indications for
$D > 1$ will be given at the end of the paper. \par

In our one-dimensional problem we shall \emph{not} assume parity
invariance (corresponding to central forces in $D = 3$) nor we
shall  exclude a zero-energy resonance. We were motivated by the
spectral density considerations of Niemi and Semenoff \cite{Nie}
for fermions in solitonic backgrounds whereas the wronskian-like
technique is adopted  from Wellner as stated. The plan of the
paper is as follows: in sect. 2 we set up the scattering problem
in $D=1$, mainly to motivate notation for direct (left to right)
and right to left (``zurdo'') scattering, unitarity of the
S-matrix, etc. In the third section we introduce the spectral
density which needs to be regularized in a box, but in the
definition of \emph{relative} spectral density the space cutoff
can (and will) be removed. As stated, Levinson\'{}s theorem will
appear clearly as a sum rule for the relative spectral density.
\par

In   section fourth we shall handle two simple examples which can
be worked out directly, namely the solitonic P\"{o}schl-Teller
potential $u(x) = - 2 \   \textrm{sech}^{2} x$ and the delta
potential $u(x) = g \ \delta(x)$; the difference of generic vs.
critical potential will be cleared up, as well as the one-half
factor already noted by Barton \cite{Bar}, and present in the
one-dimensional case (another factor should appear in two
dimensions)\cite{Che}. The next section will exhibit our general
treatment of the relative spectral density for an arbitrary local
potential $u = u(x)$ (of course, decaying at $x \rightarrow \ \pm
\ \infty$ fast enough to allow for scattering). We shall emphasize
that  the spectral density is a ``hard datum'' (i.e.
spectrum-dependent as opposed to the rest of potential parameters
or ``soft data''); the spectral density is given in terms of the
derivative of the phase of the forward amplitude, this amplitude
\emph{itself} being a hard datum as well, explaining  a result
used routinely in the KdV-like evolution equations. In section 6
we carry out the momentum integration of the density to produce
the general form of Levinson\'{}s theorem; we comment briefly on
the relation with the determinant of the S-matrix which suggests
an interpretation of the theorem as an index theorem for the bound
states. In sect. 7 we set up the procedure for an \emph{arbitrary}
dimension, thus generalizing the method of Wellner; noncentral and
critical potentials (i.e., producing a zero-energy resonance) are
easily included; finally we make some comments on non-local
potentials and add some concluding remarks.

\section{\large{Scattering in one-dimension.}}

Let

\begin{equation}\label{eq:2}
    \psi''(x) + k^2 \psi (x) = u(x) \psi (x)
\end{equation}

\noindent be the Schr\"{o}dinger equation in one dimension for a local
(hence real because hermitean) potential allowing scattering, i.e.
satisfying \cite{Dei}

\begin{equation}\label{eq:3}
    \int_{-\infty}^{\infty} (1 + x^2) |u(x)| \ dx \ < \infty
\end{equation}

\noindent for positive $E = k^2$. The $D = 1$ scattering has card
$S^o$ = two modes, \emph{direct} (incoming wave towards the
right), with the asymptotic solution:

\begin{equation}\label{eq:4}
    \psi(x) \longrightarrow  \left\{ \matrix { \exp(ikx) + b(k) \exp(-ikx)  \ \ \  \textrm{for} \ \  x
    \ll 0 \cr
  t(k)\exp(ikx) \ \   \ \ \   \textrm{for}
    \gg 0 \cr}  \right.
\end{equation}

\noindent and \emph{zurdo} scattering: the incoming wave travels
towards the left

\begin{equation}\label{eq:5}
    \tilde{\psi}(x) \longrightarrow \left\{  \matrix { \exp(-ikx) + \tilde{b}(k) \exp(ikx)  \ \ \ \textrm{for} \ \  x
    \gg 0 \cr
\tilde{t}(k)\exp(-ikx) \ \ \   \textrm{for} \ \  x
    \ll 0 \cr} \right.
\end{equation}

The amplitudes $f(k) = t(k) - 1$ and $b(k)$ give the scattering
coefficients by the relations

\begin{equation}\label{eq:6}
    \sigma_{\rightarrow} = |f(k)|^2, \ \ \sigma_{\leftarrow} = |b(k)|^2
\end{equation}

\noindent together with

\begin{equation}\label{eq:7}
    |t(k)|^2 + |b(k)|^2 = 1 \ \  \ \ \textrm{or} \ \ \  |f(k)|^2 + |b(k)|^2 \equiv
    \sigma_{tot} = - 2 \Re f(k),
\end{equation}

\noindent the \emph{optical theorem} in one dimension. The
S-matrix has two channels only

\begin{equation}\label{eq:8}
    S(k) = \left( \matrix{t(k) & \tilde{b}(k)\cr b(k) &
    \tilde{t}(k)} \right)
\end{equation}

Unitarity $S^{\dag} S = S S^{\dag}$ leads to the important
relations ( with $a = |a| \exp(i\Phi_{a})$ for any amplitude)

\begin{equation}\label{eq:9}
|t(k)|^2 + |b(k)|^2 = |t(k)|^2 + |\tilde{b}(k)|^2 =
|\tilde{t}(k)|^2 + |\tilde{b}(k)|^2 = 1
\end{equation}

\noindent and

\begin{equation}\label{eq:10}
    t(k) b(k)^{*} + \tilde{b}(k) \tilde{t}(k)^{*} = 0, \ \
    \Phi[t(k)] + \Phi[\tilde{t}(k)] = \pi + \Phi[b(k)] + \Phi[\tilde{b}(k)]
\end{equation}

Now the potential being local (and hermitean of course) is real,
so \emph{time reversal} holds; it follows that

\begin{equation}\label{eq:11}
    t(k) = \tilde{t}(k)
\end{equation}

\noindent whereas an even potential would imply similarly $b(k) =
\tilde{b}(k)$, which we do  \emph{not} assume. To see
(\ref{eq:11}), notice $t(k)$ is defined as the \emph{transition}
from $\vec{k}_{in}$ to $\vec{k}_{out}$. Time reversal changes
$\vec{k}$ to $- \vec{k}$ and \emph{in} to \emph{out}; hence $t(k)$
goes to $\tilde{t}(k)$. Also we shall take $ k \geq 0$ for direct
scattering and $k \leq 0$ for \emph{zurdo} scattering; indeed then
$\tilde{t}(k) = - t(-k)$ for the above reasons. \par

We remind that we also use the terms \emph{hard data} for spectral
data, namely the spectral density, and \emph{soft data} for
orientation data, to wit, norming constants for bound states and
the phase of $b(k)$.

\section{\large{The Spectral Density.}}

We recall first elementary properties of matrices.
\emph{Completeness} for a \emph{diagonalizable} finite matrix $M$
means

\begin{equation}\label{eq:12}
    M = \sum_{m} m P_m  \ \      \textrm{or}  \  \
     \ \ 1| = \sum_{m}  P_m
\end{equation}

\noindent for eigenvalues $m$ and projectors $P_m$; the second
relation is called \emph{resolution of the identity}. In his work
on Quantum Mechanics, von Neumann \cite{vN} extended the classical
work of Hilbert on integral equations for hermitean \emph{unbound}
operators with \emph{continuous spectrum}: he called
\emph{hypermaximal} (today ``self-adjoint'') those hermitean
operators $H$ which still support a \emph{resolution of the
identity}. The resolution reads

\begin{equation}\label{eq:13}
1|_{\cal{H}} = \sum_{j=1}^{N} P_j + \int_{0}^{\infty} dP(\mu)
\end{equation}

\noindent where $\cal{H}$ is the Hilbert space of states, $1|=
1|_{\cal{H}}$ is the unit operator, $P_j$ projects to the
\emph{finite or infinite} number $(N = 0,1,...)$ of \emph{bound}
states and the continuum integral, supposed by simplicity extended
from $0$ to $\infty$, as in a standard potential problem, means
\emph{projection-valued measures}. Notice the unit operator
$1|_{\cal{H}}$ is bounded but \emph{not} in the trace class. \par

The simplest approach to the Levinson\'{}s theorem is to state the
same resolution for the free system $H_o$

\begin{equation}\label{eq:14}
    1|_{\cal{H}} = \int_{0}^{\infty} dP^{(o)}(\mu)
\end{equation}

Substracting in (\ref{eq:13}) and then taking traces we get,
supposing short-range potentials which support at most
\emph{finite} number $N$ of bound states

\begin{equation}\label{eq:15}
    - N = \int_{0}^{\infty} Tr \  d[P(\mu) - P^{(o)}(\mu)]
\end{equation}

\noindent which is, really, the most general (but rather useless)
form of the theorem. The idea is now to trade the projectors for
scattering amplitudes (or phase shifts). We know of course the
\emph{normalized} free continuum wavefunctions

\begin{equation}\label{eq:16}
    \psi_{k}^{(o)}(x) = \frac{1}{\sqrt{2\pi}} \exp(ikx).
\end{equation}

Let us normalize the continuum wavefunctions to $1$, instead to
$\sqrt{2\pi}$. Then (\ref{eq:15}) for our case really means

\begin{equation}\label{eq:17}
    - 2 \pi N = \int_{-\infty}^{\infty} dk \
    \int_{-\infty}^{\infty} dx \  [|\psi_{k}(x)|^2 - 1]
\end{equation}

\noindent as $dP/dk \sim |k><k|$, taking the traces in $x$-spaces,
defining $\psi_k(x): = <x|k>$ and using the double degeneracy $E =
(\pm k)^2$ to extend the integral to $\pm \infty$. Expression
(\ref{eq:17}) will be our fundamental formula; the $k$-integration
will be clarified below. \par

The main concern is to express (\ref{eq:17}) in terms of the
asymptotic data (phase shifts). Define the \emph{relative}
spectral density for the problem as

\begin{equation}\label{eq:18}
    \rho_{rel}(k): = \int_{-\infty}^{\infty} \ dx \ [|\psi_{k}(x)|^2 - 1]
\end{equation}

Notice the individual spectral densities diverge, i.e.
$\rho_{free} (k) = \int 1 \ dx = \infty$; only \emph{relative
densities} make sense. The idea of the proof is to relate the
spectral densities to scattering data; as we know that the
integral of the spectral density gives Levinson\'{}s, which
express the number $N$ of bound states in terms of the range of
the phase shift  ( $0 \rightarrow \infty$), (see (\ref{eq:1})), we
expect the spectral densities to be given in terms of derivatives
of the phase shifts; we shall see that this is so. Notice also the
relative spectral density is a \emph{measure}, that is, something
under an integral symbol; hence some apparent pathologies like
delta-type behaviour or box-normalization and removal are
perfectly legitimate, and not only ``heuristic'' as some authors
state.

\section{\large{Two simple examples: P\"{o}schl-Teller Soliton and Delta potentials.}}

As a warming-up exercise, let us compute the spectral densities in
two simple cases in which the exact continuum wavefunctions are
known. When considering the P\"{o}schl-Teller potential (the standard
solitonic potential) given by

\begin{equation}\label{eq:19}
    u(x) = - \frac{2}{\cosh^2 \ x} \ ,
\end{equation}

\noindent there is a \emph{single bona fide} bound state with
energy $ E = - 1$ and continuum states with \emph{no} reflection
since the potential is transparent. Indeed, the whole $ E > 0 $
wavefunction $ \psi_k(x) $ is obtained from the $u(x) = 0$ case by
Darboux\'{}s method \cite{Ble}. If $D \equiv d/dx$

\begin{equation}\label{eq:20}
\psi_k(x) \sim (D - \tanh \ x) \exp(ikx)
\end{equation}

With the correct normalization  included (so $\psi_k(x \ll 0) \sim
\exp(ikx)$),

\begin{equation}\label{eq:21}
    \psi_k(x) = \frac{(ik - \tanh \ x)}{i k + 1} \exp(ikx)
\end{equation}

The relative spectral density is therefore

\begin{equation}\label{eq:22}
    \rho_{rel}(k) = \int_{-\infty}^{\infty} [|\psi_k(x)|^2 - 1] \
    dx = \frac{-2}{k^2 + 1}
\end{equation}

The sum rule or $k$-integration gives of course

\begin{equation}\label{eq:23}
    - 2 \pi N = \int_{-\infty}^{\infty} \rho_{rel}(k) \ dk = - 2 \pi
\end{equation}

\noindent so that $ N = 1 $ as expected. There is more to say in
this case, e.g. we find a zero-energy resonance or ``half-bound''
state which corresponds to the $k = 0$ limit, i.e.

\begin{equation}\label{eq:24}
    \psi_{k=0} = -  \tanh \ x
\end{equation}

As regards the delta potential we have

\begin{equation}\label{eq:25}
    u(x) = g \ \delta(x)
\end{equation}

\noindent where in principle we leave the \emph{sign of g open}.
As the support of the potential is a point, \{0\}, the solution
for $x \neq 0$ is always asymptotic; there is also no odd wave.
For the spectral density we compute in this case

\begin{equation}\label{eq:26}
    \rho_{rel}(k) = \int_{-\infty}^{0} [|\psi_k(x)|^2 - 1] dx +
    \int_{0}^{\infty} [|\psi_k(x)|^2 - 1] dx
\end{equation}

\noindent so that

\begin{equation}\label{eq:27}
\int_{0}^{\infty} [|\psi_k(x)|^2 - 1] dx = \int_{0}^{\infty}
[|t(k)|^2 - 1] dx = - |b(k)|^2 L
\end{equation}

 \noindent for $L \rightarrow \infty$. This divergence is in fact
 \emph{spurious} and cancelled with the $x < 0$ contribution, i.e.

\begin{equation}\label{eq:28}
\int_{-\infty}^{0} [|\psi_k(x)|^2 - 1] dx =  \int_{-\infty}^{0} [
1 + |b(k)|^2 + 2 \Re \{ b(k) \exp(-2ikx) \}  - 1 ] \ dx
\end{equation}

Now we define

\begin{equation}\label{eq:29}
    2 \int_{-\infty}^{0} \Re \{ b(k) \exp(-2ikx) \} \ dx \equiv A + B
\end{equation}

\noindent so that

\begin{equation}\label{eq:30}
    A = 2 \int_{-\infty}^{0} \Re \{ b(k) \} \cos \ 2kx \ dx
\end{equation}

If we bear in mind that

\begin{equation}\label{eq:31}
    2 \int_{-\infty}^{0} \cos \ 2kx \ dx = \int_{-\infty}^{\infty}
    \cos \ 2 k x \ dx = \int_{-\infty}^{\infty} \exp(2ikx) \ dx =
    2 \pi \delta(2k) = \pi \delta(k)
\end{equation}

\noindent then we have that

\begin{equation}\label{eq:32}
    A = \Re \{ b(0) \} \pi \delta(k)
\end{equation}
 It is the case that

 \begin{equation}\label{eq:33}
    b(0) = - 1
\end{equation}

\noindent for a generic potential, including the delta, because
$\psi_{k=0} (x) = 0$ so

\begin{equation}\label{eq:34}
\exp(ikx) + b(k) \exp(-ikx) = 0 \ \ \ \textrm{as} \ \ k
\rightarrow 0
\end{equation}

\noindent and therefore $b(0) = - 1$. The exception
(\emph{critical potentials}) occurs for a zero-energy resonance,
see later, when $b(0) = 0$. On the other hand ($ L \rightarrow +
\infty$)

\begin{equation}\label{eq:35}
    \frac{B}{\Im \{ b(k) \}} = 2 \int_{-L}^{0} \sin( -2kx) \   dx
    = \frac{1 - \cos 2kL}{k}
\end{equation}

\noindent so that

\begin{equation}\label{eq:36}
    B = \frac{\Im \{b(k)\}}{k}
\end{equation}

\noindent as the oscillatory part $\cos 2kL $ gives no
contribution \emph{as a measure} when $ L \rightarrow \infty$. Now
for the delta potential itself we have (no odd wave)

\begin{equation}\label{eq:37}
    f(k) = t(k) - 1 = b(k) = \frac{g}{2ik - g}
\end{equation}

\noindent thus confirming $b(0) = - 1$. So the relative spectral
density as a whole is

\begin{equation}\label{eq:38}
\rho_{rel}(k) = - \pi \delta(k) + \frac{2g}{g^2 + 4 k^2}
\end{equation}

Notice the delta piece, which will persist for any generic
potential. Also, the dependence $\rho_{rel}(k) \propto 1/k^2$ for
$ k \gg 0 $ is general as the phase shift itself will fall with
$1/k$ (validity of the Born approximation) and we expect $
\rho_{rel}(k) \propto \delta'(k)$. Now a $k$-integration would
yield Levinson\'{}s theorem; taking care to isolate the $sign(g)$
piece we get

\begin{equation}\label{eq:39}
    - 2 \pi N = - \pi + sign(g) \  \pi
\end{equation}

In other words

\begin{equation}\label{eq:40}
    N = \frac{1 - sign(g)}{2}
\end{equation}

\noindent which is obviously correct: $ N = 1(0) \ \  \textrm{for}
\ \  g < 0 \  (g > 0)$.

\section{\large{General calculation.}}

Now we carry out the \emph{general calculation}. Starting from

\begin{equation}\label{eq:41}
    \psi''(x) + k^2 \psi (x) = u(x) \psi (x)
\end{equation}

\noindent we differentiate with respect to $k$ (represented by the
dot symbol), i.e. \cite{Wel}, \cite{Kie}

\begin{equation}\label{eq:42}
    \dot{\psi}''(x) + 2 k \psi +  k^2 \dot{\psi} (x) = u(x) \dot{\psi}(x)
\end{equation}

Why this unusual derivative? Because we expect the spectral
density to depend on $k$-derivatives of the scattering amplitudes
(or phases shift) as said before. Next we take real and imaginary
parts of the wavefunction

\begin{equation}\label{43}
    \psi_{k}(x) = \Re\{\psi_{k}(x)\} + \Im\{\psi_{k}(x)\}
\end{equation}

\noindent since,  the Schr\"{o}dinger operator being real,  each works
separately. Now we construct the wronskian for, first,
$\Re\{\psi_{k}(x)\}:= \Phi(x)$, which satisfies

\begin{equation}\label{eq:44}
    \dot{\Phi}''(x) + 2 k \Phi(x) +  k^2 \dot{\Phi} (x) = u(x) \dot{\Phi}(x)
\end{equation}

If we multiply and substract in the usual way (e.g. to get the
current) it is the case that

\begin{equation}\label{eq:45}
   [\dot{\Phi}(x) \Phi'(x) - \Phi(x) \dot{\Phi}'(x) ]' = 2 k
   \Phi^2 (x)
\end{equation}

\noindent or ($ L \rightarrow \infty$ eventually)

\begin{equation}\label{eq:46}
   \Re\{I_k\} := \dot{\Phi}(x) \Phi'(x) - \Phi(x) \dot{\Phi}'(x) |_{-L}^{L} = 2 k
   \int_{-L}^{L} \Phi^2 (x) \ dx
\end{equation}

\noindent which, together with the imaginary contribution, is the
crucial result since it allows us to express the spectral density
in terms of the asymptotic data. Next we define

\begin{equation}\label{eq:47}
    A := \Re\{I_k\} \ \   \ \   \textrm{at} \ \  L, \ \  B := \Im\{I_k\} \ \  \textrm{at} \ \  L
\end{equation}

\begin{equation}\label{eq:48}
    \ \ C := \Re\{I_k\} \ \  \textrm{at} \ \  -
    L, \ \ D := \Im\{I_k\} \ \  \textrm{at} \ \  - L
\end{equation}

\noindent and

\begin{equation}\label{eq:49}
    t(k) = |t(k)| \exp(i\varphi_t), \ \ b(k) = |b(k)| \exp(i\varphi_r)
\end{equation}

\noindent with $\Delta: = \varphi_t + k L$. In doing so

\begin{equation}\label{eq:50}
    A = |t(k)|^2  \ [ k (\dot{\varphi}_t + L ) + \cos \ \Delta \ \sin
    \ \Delta ]
\end{equation}

\begin{equation}\label{eq:51}
    B = |t(k)|^2  \ [ k (\dot{\varphi}_t + L ) - \cos \ \Delta \ \sin
    \ \Delta ]
\end{equation}

So the total forward contribution is

\begin{equation}\label{eq:52}
    A + B = 2 k  \ |t(k)|^2 \ (\dot{\varphi}_t + L)
\end{equation}

This is very nice: the factor $2 k$ of (\ref{eq:45}) appears, as
well as the derivative of the forward phase, (a hard datum; see
below) while the $L$ divergence will be spurious. \par

The calculation of the backward part is more involved as the
wavefunction is

\begin{equation}\label{eq:53}
    \psi(x \ll 0 ) = \exp(ikx) + b(k) \exp(-ikx)
\end{equation}

By repeating the former method, now for the total backward
contribution, we get (after some cancellations between $C$ and
$D$)

\begin{equation}\label{eq:54}
    C + D = 4 k L - 2 |b(k)|^2 k (\dot{\varphi}_r - L) + 2 b(k) \sin (\varphi_r +
    2 k L)
\end{equation}

As regards the regularized spectral density $\rho_L (k)$ we have
finally $A + B - C - D$ or

\begin{equation}\label{eq:55}
    2 k \rho_L (k) = 4 k L + 2 k \dot{\varphi}_t + 2 k |b(k)|^2
    (\dot{\varphi}_r - \dot{\varphi}_t) + 2 b(k) \sin (\varphi_r +
    2 k L)
\end{equation}

So the final result is

\begin{equation}\label{eq:56}
    - 2 \pi N = \int [\lim_{L \rightarrow \infty} (\rho_L (k) -
    \rho_L^{(0)}
    (k)] \ dk
\end{equation}

\noindent where as expected

\begin{equation}\label{eq:57}
\rho_L^{(0)} (k)= \int_{- L}^{L} dx = 2 L
\end{equation}

To sum up

\begin{equation}\label{eq:58}
    - 2 \pi N = \int_{- \infty}^{\infty} [\dot{\varphi}_t + |b(k)|^2
    (\dot{\varphi}_r - \dot{\varphi}_t) + \frac{b(k)}{k} \sin
    (\varphi_r+ 2 k L) ] \ dk.
\end{equation}

As a matter of fact the first term of the integrand would be the
\emph{relative spectral density}. What about the back phase and
the oscillatory third term? We find that, first,

\begin{equation}\label{eq:59}
    \int_{- \infty}^{\infty} \frac{b(k)}{k} \sin
    (\varphi_r+ 2 k L) \ dk = \pi b(0)
\end{equation}

\noindent as shown in the Appendix. So the $\sin
    (\varphi_r+ 2 k L)/k$ integral is really $\pi \delta(k)$ as a
distribution. This is completely rigorous because we are talking
of measures (projection-valued measures) and the delta is itself a
measure (not so the delta prime). \par

As regards the second term in (\ref{eq:58}) we write the
\emph{zurdo} contribution with $k > 0$ and integrate then from $0$
to $\infty$, i.e.

\begin{equation}\label{eq:60}
    |b(k)|^2 (\dot{\varphi}_r - \dot{\varphi}_t) + |\tilde{b}(k)|^2
    (\dot{\tilde{\varphi}}_r  - \dot{\varphi}_t) = |b(k)|^2 (
    \dot{\varphi}_r  + \dot{\tilde{\varphi}}_r - 2 \dot{\varphi}_t )
    = 0
\end{equation}

\noindent because (see (\ref{eq:9}) and (\ref{eq:10})) $|b(k)| =
|\tilde{b}(k)|$ and $2 \varphi_t - \varphi_r - \tilde{\varphi}_r =
\pi $. So the final expression for the relative spectral density
is

\begin{equation}\label{eq:61}
    \rho_{rel} (k) = \dot{\varphi}_t + \pi b(0) \delta (k), \ \ k
    \geq 0
\end{equation}

\noindent see e.g. \cite{Kie}, \cite{Bia}. The first term,
derivative of the forward phase, comes by no surprise and
represents the germ of Levinson\'{}s theorem. The second one gives
a universal contribution since

\begin{equation}\label{eq:62}
    b(0) = - 1 \ \   (\textrm{generic});
     \ \ \ b(0) = 0 \ \
    (\textrm{critical})
\end{equation}

\noindent as we already discussed. We remark here how the spectral
density is  a hard function and therefore the forward phase, but
not the backward one, is a hard datum. Indeed, at least for
finite-range potentials the forward amplitude can be expressed in
terms of the bound states plus an integral over the modulus of the
reflected amplitudes \cite{Lam}. Again the usual proof is based on
analytic continuation, whereas ours stems directly from the
definition of hard data as spectral data. \par

We can easily compute the value of $\rho_{rel}(k)$ for large $k$.
From the Born approximation

\begin{equation}\label{eq:63}
    f(k) = t(k) - 1 \simeq \frac{1}{2ik}\int_{-\infty}^{\infty} u(x) \
    dx \equiv \frac{- i}{2 k}<u>
\end{equation}

Hence

\begin{equation}\label{eq:64}
    \tan \varphi_{t}(k) \simeq \frac{- < u >}{2k} \simeq
    \varphi_{t}(k) \simeq \frac{- < u >}{2k}
\end{equation}

\noindent or

\begin{equation}\label{eq:65}
\rho_{rel}(k) = \frac{d \varphi_t(k)}{dk} \rightarrow \frac{< u
>}{2 k^2} \ \ (k \gg 0 )
\end{equation}

\section{\large{The Sum Rule.}}

The crowning result is Levinson\'{}s theorem in one dimension:
integrating (\ref{eq:61}) from $k = 0$ to $k = \infty$ we find

\begin{equation}\label{eq:66}
    N = \frac{[\varphi_t(0) - \varphi_t(\infty)]}{\pi} - \frac{b(0)}{2}
\end{equation}

\noindent as first given (except that he took only $ b(0) = -1$)
by Barton \cite{Bar}. \par

We already showed that $b(0) = - 1$ for a generic potential, that
is, when the full wavefunction $\psi_k(x) \rightarrow 0$ for $ k
\rightarrow 0$. When the potential is critical the zero-energy
wave function is non-zero, just starting from

\begin{equation}\label{eq:67}
    \psi_{k \rightarrow 0} = [\exp(ikx) + b(k) \exp(-ikx)]_{k=0} =
    1 \ \ \ (x \ll 0)
\end{equation}

\noindent and so $b(0) = 0$ for $u(x)$ critical. Hence then
$|t(0)| = 1$, but the phase depends on the potential. In
particular if $u(x)$ is even, the zero-energy resonance is either
even or odd and therefore $t(0) = \pm 1$, $\psi_{k=0}(x)$
even/odd. For example, in the P\"{o}schl-Teller case given by

\begin{equation}\label{eq:68}
    u(x) = - \frac{\ell (\ell + 1 )}{\cosh^2 \ x}, \ \ \ \   \ell \ \
    \textrm{integer}
\end{equation}

\noindent the zero-energy resonance has the parity of $\ell$. Thus
 the first member $\ell = 1$ of the series  obtains (see
 (\ref{eq:24}))

\begin{equation}\label{eq:69}
    \psi_{k=0}(x) = - \tanh \ x.
\end{equation}

Now we come to the most conspicuous aspect of the one-dimensional
scattering, namely the one-half factor in the generic case
(\ref{eq:66}). As hinted at by Barton, the reason is related to
the fact that any attractive potential binds in one-dimension
(this theorem seems due to R. Peierls \cite{Pei}). Then, the $u(x)
= 0$ potential is critical, that is, increasing it a little bit in
the attractive side produces  a \emph{bona fide} bound state.
Indeed, $b(k) = 0$ for no potential, the earmark for a critical
potential, namely transparency at $ k = 0$. We like to call it
supercritical because it is transparent, i.e. there is no
reflection at \emph{any} energy.
\par

The phase of the forward amplitude is related to the determinant
of the S-matrix. From (\ref{eq:8}) and the phase relation
(\ref{eq:10})

\begin{equation}\label{eq:70}
    \textrm{Det} \  S(k) = t \tilde{t} - b \tilde{b} = t^2 - |b|^2
    \exp i(\varphi_r + \tilde{\varphi}_r) = |t|^2 \exp(2i\varphi_t)
    - e^{i \pi} |b|^2 \exp(2i\varphi_t) = \exp(2 i \varphi_t)
\end{equation}

\noindent already noticed by many people, e.g. \cite{Jau}. \par

Therefore, if $s(k):= \textrm{Det} S(k)$,

\begin{equation}\label{eq:71}
    [N + b(0)/2] = \frac{- 1}{2 \pi i} \int \frac{\dot{s}}{s} \ dk
\end{equation}

\noindent very similar to the conventional proof, as the integral
can be performed in the complex plane \cite{Lv1}. \par

The $1/2$ contributions for critical potentials, both free $u = 0$
and interacting, are reminiscent of the $\eta$-invariant in the
APS index theorem for manifolds with boundary; indeed, this can be
seen explicitly in the supersymmetric formulation, in which it
appears as the index of the Dirac operator, giving e.g. fermion
numbers $1/2$ (see \cite{Nie} and \cite{Far}). Moreover, the $1/2$
value is characteristic of time-reversal invariant systems, both
here and in the fractionization case.

\section{\large{Final remarks.}}

In principle our method can be applied in arbitrary dimension $D$:
the Schr\"{o}dinger equation for scattering reads

\begin{equation}\label{eq:72}
    \nabla^2 \psi(\vec{r}) + k^2 \ \psi(\vec{r}) = u(\vec{r}) \ \psi(\vec{r})
\end{equation}

\noindent where $\vec{r} \in R^D$. Again $u(\vec{r})$ is real so
taking real and imaginary parts and differentiating with respect
to $k$, we can get e.g. for the real part $\Phi = \Re\{\psi\}$

\begin{equation}\label{eq:73}
    \int_{\partial V} [\dot{\Phi} \nabla \Phi - \Phi \nabla
    \dot{\Phi}] \ d^{D-1}\sigma = 2k \int_V \Phi^2 \ d^D \vec{r}
\end{equation}

The first term is evaluated asymptotically in terms of

\begin{equation}\label{eq:74}
    \psi_k (r) \rightarrow \exp(i k r \cos \ \theta) + r^{-(D-1)/2}
    f(\Omega) \exp(ikr)
\end{equation}

\noindent as $r \gg 0$, without supposing $u(\vec{r}) =
u(|\vec{r}|)$, i.e. non-central potentials are included. The
second term in (\ref{eq:73}) is related, as before, to the
spectral density. The calculation proceeds in the same way, except
that for $D > 2$ the case  $u(\vec{r}) = 0$ is not critical. We
refrain to reproduce the well known results  both for $D = 2$ and
for $ D > 2$. (For $D = 2$ see \cite{Che}, \cite{Nw1}, \cite{Nw2})
\par

Non-local potentials require a different strategy, because then
time reversal $T$ does not necessarily hold. Here we want just to
show how in one-dimension a non-local \emph{real} potential, which
is $T$ invariant, gives, in general, the same result as the local
case. The Schr\"{o}dinger equation is now

\begin{equation}\label{eq:75}
    \psi''(x) + k^2 \psi (x) = \int_{- \infty}^{\infty} u(x,y) \psi (y) \ dy
\end{equation}

\noindent where

\begin{equation}\label{eq:76}
    u(x,y) = u(y,x)^*
\end{equation}

\noindent in all cases from hermiticity of the hamiltonian $H$. If
moreover $T$ reversal holds, $u(x,y)$ is real (hence symmetric)
and we get from (\ref{eq:72})

\begin{equation}\label{eq:77}
    \dot{\psi}''(x) + k^2 \dot{\psi}(x) + 2k \ \psi = \int_{- \infty}^{\infty} u(x,y) \dot{\psi} (y) \ dy
\end{equation}

Once again it suffices to take real and imaginary parts and write
the wronskian to eliminate the term of the potential, so
Levinson\'{}s theorem seems to hold untouched. Ho\-we\-ver, for a
non-local potential there might be exceptionally bound states
embeded in the continuum; for the form of the theorem in these
cases see \cite{Mar}. \par

As a final comment we want to compare the optical theorem in $D$
dimensions \cite{Boy} with this Levinson\'{}s theorem. Both depend
on hard data, hence the appearance of the forward amplitude is to
be expected. Also, they  are  interference-type formulae, linear
on $t$, and represent the same completeness. For the optical
theorem it is in coordinate space and takes the form of a
conserved current, indeed the Noether current associated to the
classical lagrangian reproducing the time-dependent Schr\"{o}dinger
equation with a global phase invariance. For the spectral density
the completeness appears in $k$-space; the sum rule for this case
is a kind of global invariant of the problem. \par

 The generalization of this presentation for the arbitrary $D$
dimensional case is in progress \footnote{In collaboration with M.
Aguado.}.

\centerline{\bf{\large{Appendix}}} \bigskip

Equation (\ref{eq:59}) is

\begin{equation}\label{eq:78}
    I \equiv \int_{- \infty}^{\infty} \frac{b(k)}{k} \ \sin (
    \varphi_r + 2 k L) \ dk \  \  \  \textrm{for} \ \  L \rightarrow \infty
\end{equation}

Define $2 k L = k'$; then

\begin{equation}\label{eq:79}
    I = \int_{- \infty}^{\infty} b(k'/2 L) \ \sin[\varphi_r (k'/2
    L) + k' ] \ dk'/k' = b(0) \int_{- \infty}^{\infty} \sin(k') \
    dk'/k' = \pi b(0)
\end{equation}

\noindent as both $\varphi_r(0)$, $\tilde{\varphi}_r(0)$ and
$b(0)$ are regular.
\bigskip

{\bf{ Acknowledgements.-}} One of us (LJB) thanks Profs. E.C.G.
Sudarshan (Austin) and A. Galindo (Madrid) for some early
discussions. Both of us have profited from comments by our
colleagues M. Asorey, J.G. Esteve, M. Aguado and A. Segu\'{\i}. Partial
support by grants from M.C. y T. (Spain): FPA-2003-02948 (LJB) and
BFM-2003-01300 (JC) are grafetully acknowledged.

\vfill \eject

\end{document}